\documentclass[epsf,usegraphicx]{mn2e}

\title[A CV with an orbital period of 56.6 min?]{BOKS 45906: a CV with
  an orbital period of 56.6 min in the {\kep} field?}

\author[]
{Gavin Ramsay$^{1}$, Steve B. Howell$^{2,10}$, 
Matt A. Wood$^{3}$, Alan Smale$^{4}$, Thomas Barclay$^{2,5}$, \and Sally A. Seebode$^{6,10}$, 
Dawn Gelino$^{7,10}$, Martin Still$^{2,4}$, John K. Cannizzo$^{8,9}$ \and \\
$^{1}$Armagh Observatory, College Hill, Armagh, BT61 9DG, UK\\
$^{2}$NASA Ames Research Center, Moffett Field, CA 94095, USA\\
$^{3}$Physics \& Astronomy Department, Texas A\&M University-Commerce, 
Commerce, TX 75429, USA\\
$^{4}$NASA/Goddard Space Flight Center, Greenbelt, MD 20771, USA\\
$^{5}$Bay Area Environmental Research Institute, Inc., 560
Third St. West, Sonoma, CA 95476, USA\\
$^{6}$San Mateo High School, San Mateo, CA 94401, USA\\
$^{7}$NASA Exoplanet Science Institute, Caltech, Pasadena, CA 91125, USA\\
$^{8}$CRESST and Astroparticle Physics Laboratory
NASA/GSFC, Greenbelt, MD 20771, USA\\
$^{9}$Department of Physics, University of Maryland,
Baltimore County, 1000 Hilltop Circle, Baltimore, MD 21250, USA\\
$^{10}$Visiting Astronomer, Mt. Palomar Observatory\\
}

\date{Accepted 2013 November 17.  Received 2013 November 17; in original form 2013
October 29}

\begin{document}
\newcommand{\Msun} {$M_{\odot}$}
\newcommand{\kep}{\it Kepler}
\newcommand{\swift}{\it Swift}
\newcommand{\Porb}{P_{\rm orb}}
\newcommand{\nuorb}{\nu_{\rm orb}}
\newcommand{\eplus}{\epsilon_+}
\newcommand{\eminus}{\epsilon_-}
\newcommand{\cd}{{\rm\ c\ d^{-1}}}
\newcommand{\MdotL}{\dot M_{\rm L1}}
\newcommand{\Ldisk}{L_{\rm disk}}
\newcommand{\src}{BOKS 45906}
\newcommand{\ergscm} {ergs s$^{-1}$ cm$^{-2}$}

\maketitle

\begin{abstract}

{\src} was found to be a blue source in the
Burrell-Optical-Kepler-Survey which showed a 3 mag outburst lasting
$\sim$5 d. We present the {\kep} light curve of this source which
covers nearly 3 years. We find that it is in a faint optical state for
approximately half the time and shows a series of outbursts separated
by distinct dips in flux. Using data with 1 min sampling, we find
clear evidence that in its low state {\src} shows a flux variability
on a period of 56.5574$\pm$0.0014 min and a semi-amplitude of $\sim$3
percent. Since we can phase all the 1 min cadence data on a common
ephemeris using this period, it is probable that 56.56 min is the
binary orbital period. Optical spectra of {\src} show the presence of
Balmer lines in emission indicating it is not an AM CVn (pure Helium)
binary. {\sl Swift} data show that it is a weak X-ray source and is
weakly detected in the bluest of the UVOT filters. We conclude that
{\src} is a cataclysmic variable with a period shorter than the
`period-bounce' systems and therefore {\src} could be the first
helium-rich cataclysmic variable detected in the {\kep} field.

\end{abstract}


\section{Introduction}

Cataclysmic Variables (CVs) are binary systems which contain a white
dwarf primary star that accretes material from a Roche lobe-filling
late-type main sequence secondary star. The secondary star's surface
contacts its Roche lobe through angular momentum loss and initiates
ballistic mass transfer to the white dwarf through the inner Lagrange
point $L_{1}$. The mass transfer that occurs at this point is mostly
stable and causes the `{\sl cataclysmic}' phenomenon.  If the white
dwarf retains enough of the accreted material, it can exceed the
Chandrasekhar limit and become a Type Ia supernova. Thus, CVs have a
broad importance for studying the physics of the accretion process in
detail, in constraining models of stellar evolution, binary evolution,
and the chemical enrichment of the Galaxy.

Although CVs have been observed for more than 100 years (e.g. Cannizzo
2012), {\kep} provided essentially uninterrupted observations of
sources for months or years which allowed the outburst properties of
CVs to be studied in a way not previously possible. Since the launch
of {\kep}, we have pursued a programme to study a range of CVs in the
{\kep} field of view (there are several dozen, see Howell et al. 2013
and Scaringi et al. 2013 for details). A brief overview of our work
can be found in Ramsay et al. (2012). Other groups which have
published recent work on {\kep} CVs include Scaringi, Groot \& Still
(2013), Kato \& Maehara (2013), Gies et al. (2013) and Osaki \& Kato
(2013).

One source in the {\kep} field, {\src} (KIC 9778689, $\alpha=19^{\rm
  h}40^{\rm m}16.2^{\rm s}$ $\delta = +46\degr 32^{'}47.9^{''}$
J2000.0, taken from the Kepler INT Survey), was identified as a blue
source in the Burrell-Optical-Kepler-Survey (BOKS) which was a
pre-launch survey of the central region of the {\kep} field of
view. {\src} was seen to undergo what appeared to be a typical dwarf
nova outburst, rising from a quiescent magnitude of $R$=20 to 16.5 and
lasting for about 5 d. A subsequent study by Howell (unpublished) over
the past few years has shown {\src} to reside near $V$=22.5 and no
further outbursts such as detected by Feldmeier et al. (2011) have
been noted. There is a faint X-ray source (0.02 ct/s) detected in {\sl
  ROSAT} All-Sky Faint Survey Catalogue (Voges et al. 1999) which is
3.5$^{''}$ distant from the optical position of {\src}.

As part of a broader programme to study CVs (and accreting objects in
general) in NASA's {\kep} field of view, we added BOKS 45906 to our
{\kep} programme.  In this paper we present the {\kep} data of {\src}
with additional data obtained using the Isaac Newton Telescope (INT),
{\sl Swift}, and the Hale 200$^{''}$ Telescope, and discuss its likely
nature.

\section{Optical Colours}

\begin{figure}
\begin{center}
\setlength{\unitlength}{1cm}
\begin{picture}(6,12)
\put(-1.5,-1.5){\includegraphics{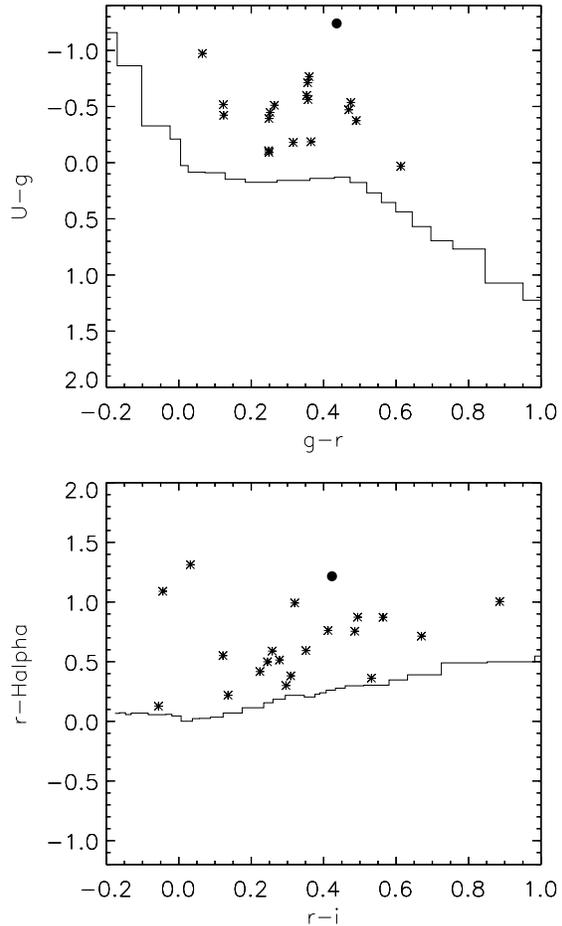}}
\end{picture}
\end{center}
\caption{The colours of {\src} (the filled circle) in the $U-g, g-r$
  (top panel) and the $r-i, r-$H$_{\alpha}$ (lower panel). We show the
  track of the unreddeneded main sequence taken from Groot et
  al. (2009) (top) and Drew et al. (2005) (bottom). We also show the
  colours of known CVs in the {\kep} field (taken from Scaringi et
  al. 2013 and Howell et al.  2013).}
\label{kiscolours} 
\end{figure}

The {\sl Kepler INT Survey} (KIS, Greiss et al. 2012a,b) obtained
$Ugri$H$_{\alpha}$ photometry of the vast majority of sources in the
{\sl Kepler} field. The photometry and colours of {\src} are
$g$=19.85, $U-g$=--1.24, $g-r$=0.44, $r-i$=0.42 and
$r-$H$_{\alpha}$=1.22 indicating it has colours consistent with those
of CVs (Figure \ref{kiscolours}).  The {\kep} data imply that these
KIS observations took place when the source was in a high optical
state (\S 3). Although the KIS multi-colour observations took place
within 8 min of each other, our INT observations (cf \S
\ref{int-phot}) indicate that some degree of caution should be taken
since it can change brightness by up to 0.4 mag on this timescale.

\section{Kepler Photometric Observations}

\begin{table*}
\begin{center}
\begin{tabular}{lllll}
\hline
Quarter & \multicolumn{2}{c}{Start} & \multicolumn{2}{c}{End} \\
        & MJD   & UT  & MJD & UT\\ 
\hline
Q6 (LC) & 55371.947 & 2010 Jun 24 22:46 & 55461.794 & 2010 Sep 22 19:04 \\
Q7 (LC) & 55462.673 & 2010 Sep 23 16:10 & 55552.049 & 2010 Dec 22 01:09 \\
Q8 (LC) & 55567.865 & 2011 Jan 06 20:45 & 55634.846 & 2011 Mar 14 20:18 \\
Q9 (LC) & 55641.017 & 2011 Mar 21 00:24 & 55738.424  & 2011 Jun 26 10:10 \\
Q10 (LC) & 55739.343 & 2011 Jun 27 08:16 & 55832.766 & 2011 Sep 28 18:24 \\
Q11 (LC) & 55833.706 & 2011 Sep 29 16:56 & 55930.827 & 2012 Jan 04 19:50 \\
Q12 (LC) & 55931.910 & 2012 Jan 05 21:50 & 56014.523 & 2012 Mar 28 12:33\\
Q13 (LC) & 56015.238 & 2012 Mar 27 05:42 & 56105.554 & 2012 Jun 27 13:17 \\
Q14 (LC) & 56106.637 & 2012 Jun 28 15:17 & 56203.820 & 2012 Oct 03 19:40 \\
Q15 (LC) & 56205.985 & 2012 Oct 05 23:38 & 56303.638 & 2013 Jan 11 15:18 \\
Q16 (LC) & 56304.598 & 2013 Jan 12 14:21 & 56390.460 & 2013 Apr 08 11:02 \\
\hline
Q6-1 (SC) & 55371.937 &  2010 Jun 24 22:29& 55399.032 & 2010 Jun 24 22:29\\
Q6-2 (SC) & 55399.870 &  2010 Jul 22 20:53& 55430.786 & 2010 Jul 22 20:53\\
Q6-3 (SC) & 55431.685 &  2010 Aug 23 16:26& 55461.804 & 2010 Aug 23 16:26\\
Q7-1 (SC) & 55462.663 &  2010 Sep 23 15:54& 55492.781 & 2010 Sep 23 15:54\\
Q7-2 (SC) & 55493.538 &  2010 Oct 24 12:54& 55522.737 & 2010 Oct 24 12:54\\
Q7-3 (SC) & 55523.616 &  2010 Nov 23 14:47& 55552.059 & 2010 Nov 23 14:47\\
Q8-1 (SC) & 55567.855 &  2011 Jan 06 20:30& 55585.550 & 2011 Jan 06 20:30\\
Q8-2 (SC) & 55585.612 &  2011 Jan 24 14:40& 55614.708 & 2011 Jan 24 14:40\\
Q8-3 (SC) & 55614.770 &  2011 Feb 22 18:29& 55634.856 & 2011 Feb 22 18:29\\
Q11-1 (SC) & 55833.700 & 2011 Sep 29 16:42& 55864.775 & 2011 Sep 29 16:42\\
Q11-2 (SC) & 55865.531 & 2011 Oct 31 12:45& 55895.732 & 2011 Oct 31 12:45\\
Q11-3 (SC) & 55896.611 & 2011 Dec 01 14:39& 55930.837 & 2011 Dec 01 14:39\\
Q15-1 (SC) & 56205.976 & 2012 Oct 05 23:24& 56236.809 & 2012 Oct 05 23:24\\
Q15-2 (SC) & 56237.770 & 2012 Nov 06 18:29& 56267.889 & 2012 Nov 06 18:29\\
Q15-3 (SC) & 56268.727 & 2012 Dec 07 17:27& 56303.648 & 2012 Dec 07 17:27\\
\hline
\end{tabular}
\end{center}
\caption{Journal of {\kep} observations. The start and end MJD and UT
  dates are the mid point of the first and final cadence of the LC
  time series for each quarter respectively.}
\label{kepler-log}
\end{table*}

\subsection{Analysis of the {\kep} data}

The detector on board {\kep} is a shutterless photometer using 6 sec
integrations and a 0.5 s readout. There are two modes of observation:
{\it long cadence} (LC), where 270 integrations are summed for an
effective 29.4 min exposure (this includes deadtime), and {\it short
  cadence} (SC), where 9 integrations are summed for an effective 58.8
s exposure.  Gaps in the {\kep} data streams result from, for example,
90$^\circ$ spacecraft rolls every 3 months (called Quarters), and
monthly data downloads using the high gain antenna.

\begin{figure}
\begin{center}
\setlength{\unitlength}{1cm}
\begin{picture}(6,6.5)
\put(-3,-1){\includegraphics{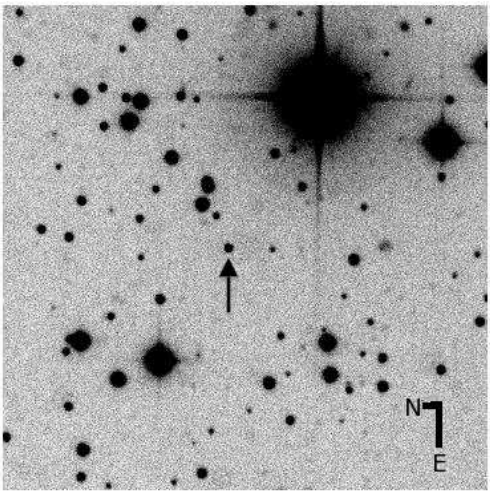}}
\end{picture}
\end{center}
\caption{The finding chart of {\src} (marked by an arrow) made using
  the INT in July 2013. It was made by combining 11 separate $g$ band
  images, the effective exposure being 660 sec. The image is 2 arcmin
  across.}
\label{intfind} 
\end{figure}

Kepler data are available in the form of FITS files which are
distributed by the Mikulski Archive for Space Telescope
(MAST)\footnote{http://archive.stsci.edu/kepler}. For LC data each
file contains one observing quarter worth of data whereas for SC data
one file is created per month.  After the raw data are corrected for
bias, shutterless readout smear, and sky background, time series are
extracted using simple aperture photometry (SAP).  The start and end
times of each quarter of {\kep} data which are used in this study are
shown in Table \ref{kepler-log}. SC mode data were obtained in
Q6--8,11 \& 15. (We note that when SC data are obtained, LC data are
also produced).  {\src} is located in a relatively crowded field
($b=11.6^{\circ}$). There is a faint ($g$=21.8) star 8.9$^{''}$ to the
NW of {\src} and a brighter ($g$=18.3) star 13.0$^{''}$ also to the NW
of {\src} (these can be seen in Figure \ref{intfind}) where the source
magnitudes come from the {\sl Kepler INT Survey} (Greiss et
al. 2012a,b). Using the {\tt PyKe}
software{\footnote{http://keplergo.arc.nasa.gov/PyKE.shtml}}, we find
that the brighter of these two stars is likely to contaminate the
standard photometric results of {\src} ({\kep} pixels are 3.98$^{''}$
square). We therefore used the {\tt PyKe} tasks {\tt keppixseries,
  kepmask} and {\tt kepextract}, to extract photometric data from one
pixel which clearly showed the presence of {\src} in all quarters of
data.

\begin{figure*}
\begin{center}
\setlength{\unitlength}{1cm}
\begin{picture}(12,9)
\put(-2,-0.5){\includegraphics{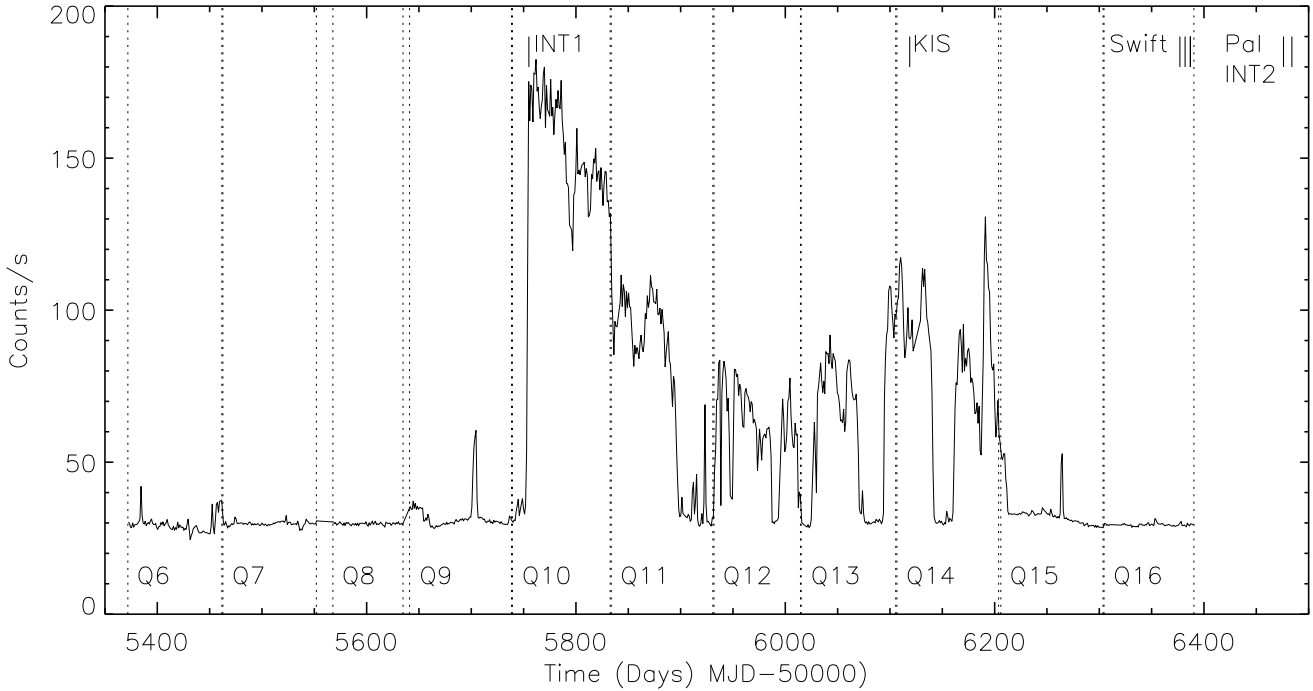}}
\end{picture}
\end{center}
\caption{The {\kep} LC {\kep} light curve of BOKS 45906
  which covers Q6--16 binned into 1 d bins. The time unit is in MJD
  - 50000.0. We mark the dates when INT, KIS, Swift and Palomar (Pal)
  observations were made.}
\label{longcadencelight} 
\end{figure*}

Data which were not flagged as `SAP\_QUALITY=0' in the {\tt FITS}
files were removed (for instance, time intervals of enhanced solar
activity). When {\src} was in a low optical state, we corrected for
systematic trends (Kinemuchi et al. 2012). (In a high optical state
this `correction' introduced spurious effects). Due to the rotation of
the spacecraft the source lies on a different chip each quarter. We
therefore applied a small correction to the resulting flux so that
there were no discrete jumps in the light curve from quarter to
quarter. We show the resulting light curve binned into 1 d means in
Figure \ref{longcadencelight}.

\subsection{Overall characteristics of the light curve}

The SC light curve for quarters Q6-Q8 was presented initially in
Howell et al. (2013). In their paper, {\src} was seen to be in a low
state with only a few low amplitude flare-like events noted.  The LC
{\kep} light curve now covers 1018 d (2.8 years). For the first year
of observation, {\src} was typically in a low state, with occasional
short duration flux enhancements. However, at MJD$\sim$55752 there is
a rapid ($\sim$2 d) increase in the flux (a factor of 5.7, or 1.9 mag)
of {\src}. Over the next $\sim$145 d the flux gradually decreases
until it makes a rapid $\sim$2--3 d descent back to its `quiescent'
state. However, it only remains in a quiescent state for $\sim$10 d
thereafter showing short duration events for the next $\sim$280 d
going into a series of bright states interspersed with `dips' in the
light curve which have sharp ingress and egrees features when the
source returns to its quiescent flux. The last 100 d of the light
curve show it in a quiescent state apart from one short duration
event. The fact that the flux returns to quiescence after the initial
long outburst (and the following outbursts) makes it different in
character to the `echo' outbursts seen in WZ Sge (eg Patterson et
al. 2002).

\subsection{A search for short period variations}

SC data which an effective exposure of 58.85 s were obtained in five
quarters, Q6, Q7, Q8, Q11 and Q15 (see Table \ref{kepler-log}). We
removed data from time intervals of enhanced Solar background in a
similar way to the LC analysis and we used the pixel level data and
extracted light curves using the one pixel that {\src} was most
strongly detected on.

\begin{figure*}
\begin{center}
\setlength{\unitlength}{1cm}
\begin{picture}(12,5)
\put(-3,-6.2){\includegraphics{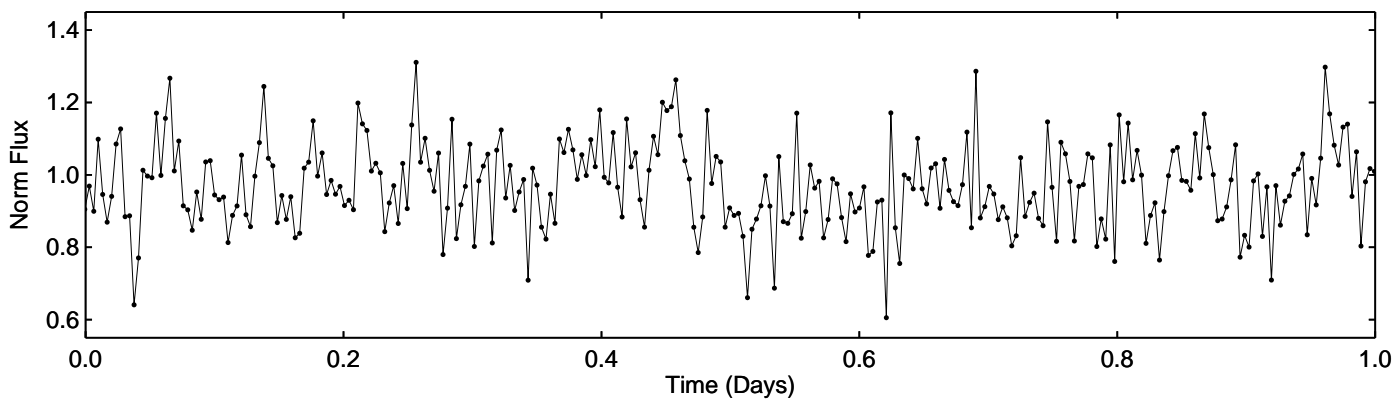}}
\end{picture}
\end{center}
\caption{The light curve of {\src} using SC data from Q6-1 where we
  have binned the data into 5 min bins and show only one d of data. The
  data have been normalised so that the mean flux is set at unity.}
\label{light-q6-1} 
\end{figure*}

\begin{figure}
\begin{center}
\setlength{\unitlength}{1cm}
\begin{picture}(6,5.5)
\put(-1.5,-0.2){\includegraphics{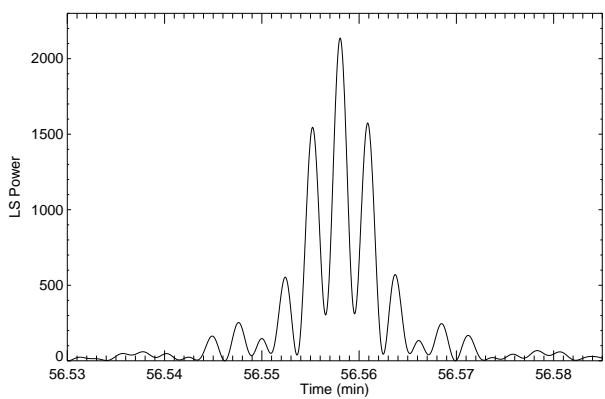}}
\end{picture}
\end{center}
\caption{The Lomb Scargle Power Spectrum of {\src} using SC data from
  Q6,7,8 and 15 when the {\src} was in a low state. There is a very
  strong peak in the power spectrum corresponding to a period of
  56.557 min.}
\label{power} 
\end{figure}

For the Q6-1 data we first removed data from the time interval where
there was a flux enhancement lasting $\sim$2 d. The mean count for
this resulting light curve (Figure \ref{light-q6-1}) was 29.6 ct/s
with a standard deviation (rms) of 6.3 ct/s (21 percent). In contrast,
the pixel corresponding to 319,991 in the {\tt keppixseries} map,
which we took to represent the background, gave a mean flux of -0.3
ct/s indicating that {\src} was clearly detected. The Lomb Scargle
Power Spectrum of the Q6-1 light curve of {\src} is shown in Figure
\ref{power} and indicates a very prominent peak at a period of 56.56
min.

Although there is no previously known spurious period caused by
instrumental effects at 56.56 min, we note it is close to twice the
effective exposure of the LC data (58.867 min). We therefore searched
for a periodic signal in all of the other pixels in the Q6 data and
found evidence for this period in only 3 other pixels which were
adjacent to the pixel where {\src} was most strong detected. In
addition we also extracted the {\tt RAW\_FLUX} data (i.e. using the
original data which have not been corrected for flat-fielding etc) and
find evidence for a period of 56.56 min in the same three pixels and
none of the others.

\begin{figure}
\begin{center}
\setlength{\unitlength}{1cm}
\begin{picture}(6,17)
\put(-2,-2){\includegraphics{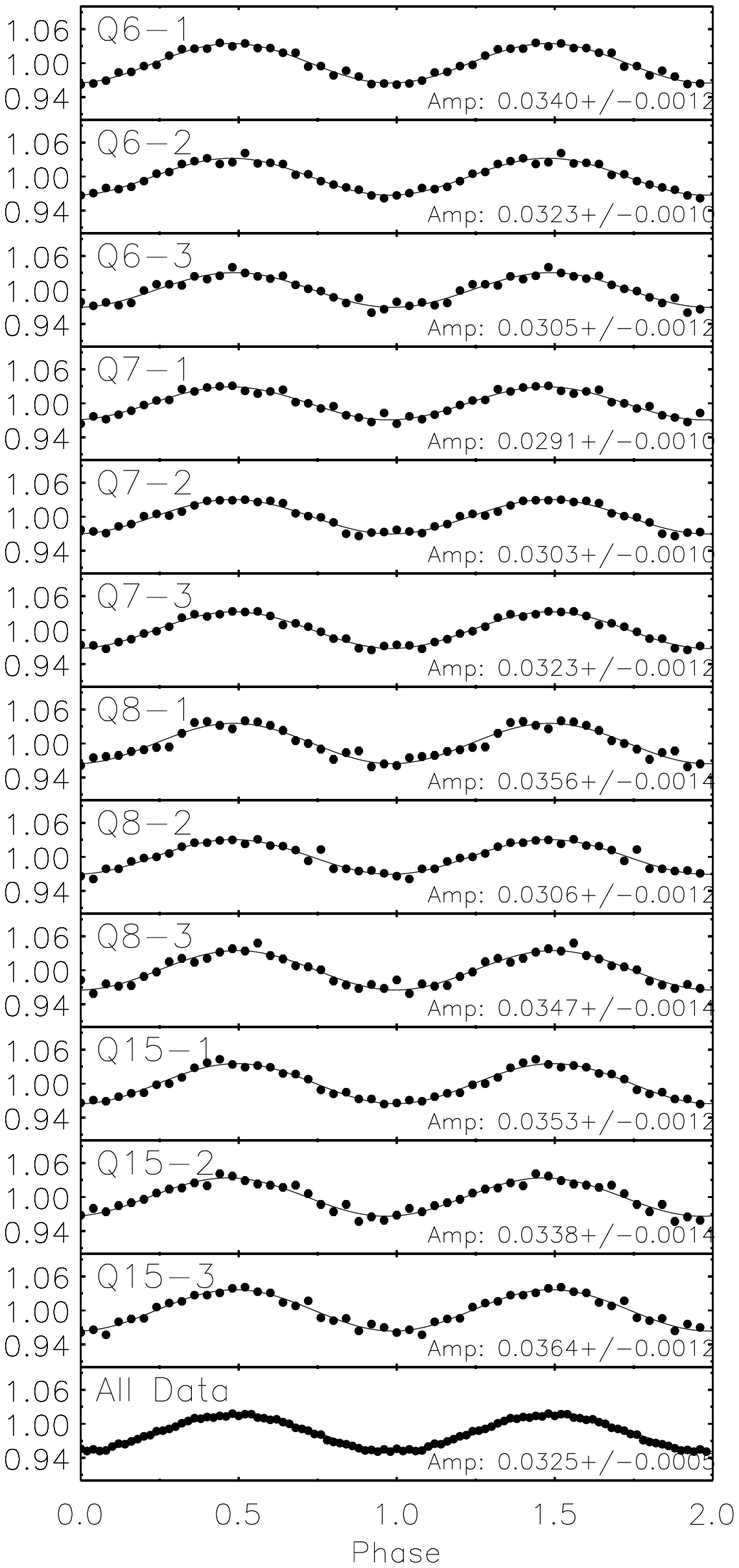}}
\end{picture}
\end{center}
\caption{The SC data of {\src} folded on a period of
  56.55804 min with phase zero being HJD 24555372.4512.}
\label{light-phase} 
\end{figure}

For all of the other SC data we extracted the light curve for {\src}
in the same manner and removed data from those time intervals where
the source was in a brighter optical state. We find each quarter of
data has a strong peak in the corresponding Lomb Scargle Power Spectra
at 56.56 min. Taking all of this data we find a period of 56.55804 min
(we show the Lomb Scargle Power spectra of all low state data in
Figure \ref{power}).  We find that the following ephemeris:

\begin{equation}
T_{o} = HJD 2455372.4512(5) + 0.039276(1) E \noindent
\end{equation}

can phase each of the low optical state datasets so that the minimum
corresponds to $\phi$=0.0 (Figure \ref{light-phase}). (The error on
the folding epoch was derived from the mean error on the phase of the
flux minimum in the folded light curves and the error on the period is
the full-width-half-maximum of the main peak in the power spectrum
shown in Figure \ref{power}). We fitted a sine wave to each of these
folded light curves and show the best-fit and error of the
half-amplitude for each epoch in Figure \ref{light-phase}. We find
that the mean half-amplitude is 3.3 percent which is much smaller than
the rms of the Q6-1 light curve (21 percent).  Given the stable nature
of this period we suggest that the period of 56.56 min represents the
orbital period of this system.

We also searched for the presence of the 56.56 min period when the
system was in a high optical state: it was detected at a much lower
significance even in the high state data. The power spectra of the
high state data is complex with peaks corresponding to periods of
$\sim$0.1--0.3 d. 

\section{Ground Based Photometry}
\label{int-phot}

\begin{figure}
\begin{center}
\setlength{\unitlength}{1cm}
\begin{picture}(6,12)
\put(-1.5,-0.8){\includegraphics{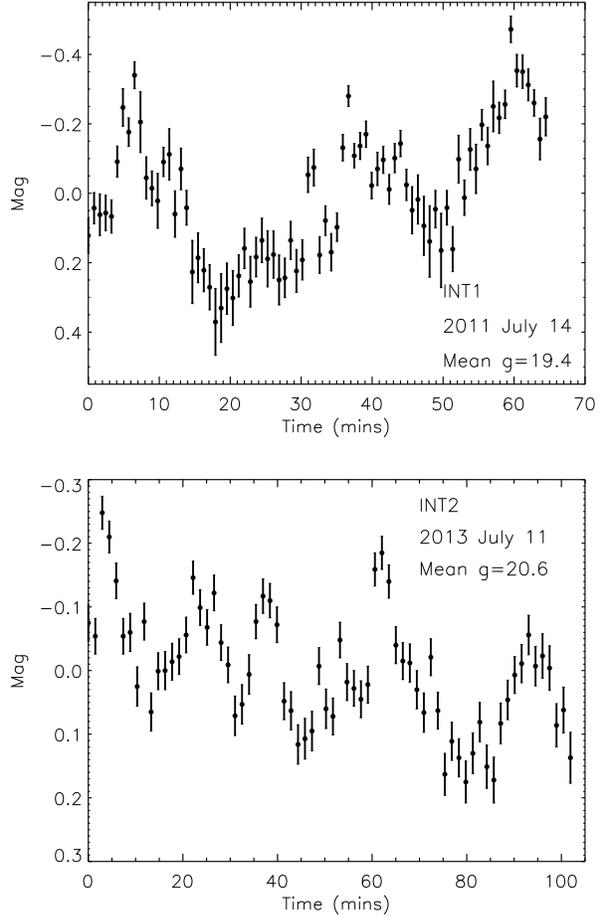}}
\end{picture}
\end{center}
\caption{Light curves of {\src} obtained using the INT in July 2011
  (top) when the source was in a bright optical state and in July 2013
  (bottom) when the source was in a faint optical state.}
\label{intlight} 
\end{figure}

We observed {\src} using the 2.5 m INT, located on the island of La
Palma, on 13 July 2011 and again on 11 July 2013 as part of the {\sl
  RATS-Survey} (Ramsay et al. 2013). The Wide Field Camera was used
with a $g$ band filter. The 2011 observations were made using a 20 sec
exposure and the light curve (Figure \ref{intlight}) covered 65 min
and the source had a mean brightness of $g$=19.4 (using the KIS to
determine the brightness of nearby comparison stars) although {\src}
reached a peak of $g$=18.8 during these observations. Although the
source is obviously variable, there is no clear periodic signal. In
contrast the 2013 observations were made when the source was in a low
state ($g$=20.6) and exposures were 60 s in duration. Although there
are clear short timescale variations (lower panel Figure
\ref{intlight}) there is no periodic signal. The standard deviation of
this light curve is 0.09 mag which is greater than the 3 percent
semi-amplitude modulation of the 56.56 min period seen in the {\kep}
SC data. It is therefore no surprise that we do not detect the 56.56
min period in these short duration light curves.

\section{Optical Spectroscopy}
\label{spectra}

We used the 200$^{''}$ Hale Telescope on Palomar Mountain, California,
and the double beam spectrograph (DBS) to obtain a small number of
optical spectra of {\src}. The DBS has an intensified CCD acquisition
camera which can be used to take timed exposures of the local field of
view which helps to find and place faint sources on the slit. We
obtained a sequence of four 20-min exposures of {\src} on 2nd July
2013 starting at 09:38 UT. Observations made using the INT (\S 4) 9
days after these observations show {\src} to be in a low but variable
optical state.

The D-55 dichroic filter was used to split light between the blue and
red arms. The blue arm used a 1200 lines/mm grating providing of
resolution of R$\sim$7700 and covered 1500\AA\ in wavelength. The red
arm used a 1200 lines/mm grating providing a resolution of
R$\sim$10000 and a wavelength coverage of 670\AA. The slit width was
set to 1 arcsec and the usual procedures of observing
spectrophotometric standard stars and arc lamps were adhered to. Red
and blue spectra were wavelength calibrated using a HeNeAr and FeAr
arc-lamp respectively. The night was clear and provided stable seeing
near 1 arcsec.  Data reduction was done using IRAF 2-D and 1-D
routines for spectroscopic data and produced a final 1-D spectrum for
each observation.

\begin{figure}
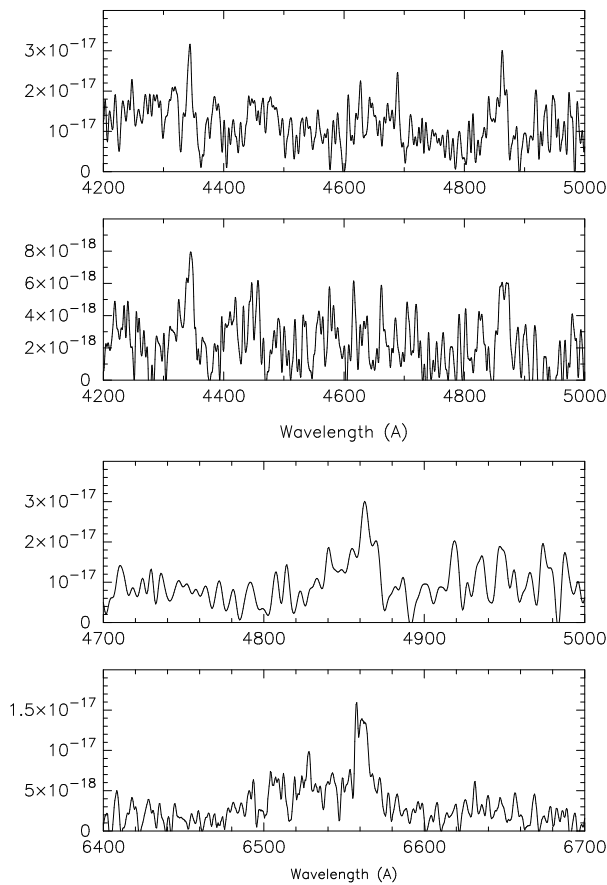

\begin{center}
\setlength{\unitlength}{1cm}
\begin{picture}(6,11.6)
\put(-2,5){\includegraphics{boks-blue-spec.ps}}
\put(-2,-1){\includegraphics{boks-red-spec.ps}}
\end{picture}
\end{center}
\caption{The upper two panels show our first and second blue spectra
  covering H$\beta$ through H$\delta$ (the flux scale is in
  erg/s/cm$^{2}$/\AA). The Balmer emission lines as well as the
  continuum level weakened in the second and subsequent exposures. The
  bottom two panels show regions centred on H$\beta$ and H$\alpha$
  from the first spectrum pair. {\src} was at minimum light during
  these observations.}
\label{spectrum} 
\end{figure}

Figure \ref{spectrum} show the Balmer lines for the first blue and red
pair as well as a comparison of the first two blue spectra.  During
the sequence of the four spectra we obtained, the star brightness
dropped by 30\% in flux from exposure 1 to 2 and continued to drop
during the remaining 2 exposures, being only 1/4 as bright during the
last exposure. Based on the slit viewer camera, BOKS 45906 was
estimated to be near $V$=22 during the observations. Due to this rapid
dimming during the four exposures, the final three red spectra and the
final two blue spectra are very low S/N, showing weak emission in a
rather noisy continuum. We cannot reliably measure the line centre or
line flux changes in the last three spectra due to the faintness of
the star.  The spectra do, however, confirm that BOKS 45906 is an
emission line source, rapidly variable, and shows a flat Balmer
decrement.

\section{Swift Observations}

We obtained target of opportunity observations of {\src} using the
NASA {\swift} satellite (Gehrels et al. 2004) at four epochs separated
by $\sim$5 d. As can be seen from Figure \ref{longcadencelight}
they were made when the source was in a low optical state.  Because
{\swift} is in a low Earth orbit, each observation sequence (which
makes up an `ObservationID') is made up of typically 2--4 separate
pointings. The exposure time of each UVOT image is generally 1200 sec
in duration, while the total exposure time of the X-ray observation in
each ObservationID is typically 3--4 ksec. Observations commenced on
2013 March 26.

\begin{table}
\begin{center}
\begin{tabular}{llrr}
\hline
ObsID & Start Date & UVOT & Flux\\
      & (MJD) & Filter & \ergscm \AA$^{-1}$\\
\hline
00032763001 & 56377.55 & UVW2 & 1.2$\pm$0.2$\times10^{-17}$ \\
00032763002 & 56382.03 & UVM2 & - \\
00032763003 & 56387.33 & UVW1 & - \\
00032763004 & 56392.52 & U & - \\
\hline
\end{tabular}
\end{center}
\caption{Details of the Swift UVOT observations where we indicate the 
ObsId; start date of the observations; which filter was being used and 
the detected flux.}
\label{swift-obs}
\end{table}

\subsection{XRT observations}

The X-ray Telescope (XRT) (Burrows et al. 2005) on-board {\swift} has
a field of view of 23.6$\times$23.6 arcmin with CCD detectors. It is
sensitive over the range 0.2-10 keV and has an effective area of
$\sim$70 cm$^{2}$ at 1 keV (for comparison the {\sl ROSAT} XRT had an
effective area of $\sim$400 cm$^{2}$ at 1 keV). We examined data taken
in `photon counting' mode and used the products derived from the
standard XRT pipeline.

Since the count rate of {\src} was low we created an image using all
the XRT data in photon counting mode. This image was input to the
HEASoft tool {\tt XIMAGE} and the routine {\tt SOSTA} (which takes
into account effects such as vignetting, exposure and the point spread
function) to determine the count rate and error at the optical
position of {\src}. The total exposure was 14.1ks and the count rate
was 6.2$\pm3.1\times10^{-4}$ cts/s giving a signal to noise of 2.0. It
was therefore detected at a low significance.

\subsection{UVOT observations}

The Ultra-Violet/Optical Telescope (UVOT) on board the {\swift}
satellite has a 30 cm primary mirror and 6 optical/UV filters (Roming
et al. 2005).  Since {\swift} operates a `filter of the day', we were
not able to pre-define the filter, but images were obtained in the U
(central wavelength 347 nm, and a full width half maximum of 79 nm),
UVW1 (251 nm, 70 nm), UVM2 filter (225 nm, 50 nm), and the UVW2
(188nm, 76 nm) filters. {\src} was detected only in the UVW2 filter
(the filter with the bluest response) and then at a very low count
rate: 0.020$\pm$0.003 ct/s which corresponds to a flux of
1.23$\pm0.21\times10^{-17}$ \ergscm \AA$^{-1}$.

\section{Discussion}
\label{discuss}

The {\kep} observations of {\src} cover a time interval of nearly 3
years during which its optical flux varies by a factor of
$\sim$6. This together with the fact that it is detected as a very
weak X-ray and UV source, and shows Balmer emission lines, all point
to the fact that {\src} is an accreting source. During the low optical
state there is a strong peak in the power spectrum corresponding to a
period of 56.56 min. Given that we can phase all of the {\kep} SC data
on a common ephemeris, this strongly suggests that 56.56 min is the
orbital period of an accreting binary system (other periods which may
be expected, such as a super-hump period, would not be expected to be
as stable).

As CVs evolve over time their orbital period decreases due to the loss
of orbital angular momentum from the system. However, at some point,
the mass of the secondary star becomes so small that hydrogen burning
in the core stops and it becomes semi-degenerate. The radius of the
secondary will thereafter start to increase and the orbital period
starts to increase again. This period `bounce' is predicted to occur
at $\sim$65–-70 min (Rappaport et al.  1982, Howell, Nelson \&
Rappaport 2001, Kolb \& Baraffe 1999), and is observed at $\sim$80
min (G\"{a}nsicke et al. 2009).

There are, however, two groups of helium rich accreting binary systems
which have orbital periods shorter than 70 min. The `AM CVn' binaries,
have orbital periods between $\sim$5--70 min and spectra devoid of
hydrogen. Given we detect Balmer lines in its spectrum, {\src} cannot
be an AM CVn system. A much smaller group of accreting binaries are
also known which show strong helium and hydrogen lines and also show
outbursts (e.g. Breedt et al. 2012, Carter et al. 2013). It is thought
that these `helium rich' binaries may evolve to become AM CVn systems
over time. Although our optical spectra are low signal-to-noise we do
not see any clear evidence for helium lines. However, other CVs below
the period minimum such as EI Psc which has an orbital period of 64
min (Thorstensen et al 2002) show relatively weak helium emission
lines, which could well be hidden in low signal-to-noise spectra.

Breedt et al. (2012) give an overview of the accreting binaries with
an orbital period less than 76 min. Of the 42 then known systems, 36
of these are AM CVn binaries and 3 are confirmed (and another 3
candidate) hydrogen accreting systems. Finding another candidate
hydrogen accreting binary with an orbital period less than 76 min is
clearly of great interest from a binary evolutionary point of
view. The next step is to obtain time resolved spectra of {\src} when
it is in a bright optical state to confirm that 56.56 min is indeed
the binary orbital period.

\section{Conclusions}

We have identified one object in the {\kep} field, {\src}, which shows
a series of outbursts which can reach $\sim$2 mag in amplitude. Using
SC data when the source is in a low state, we find evidence for a flux
variation on a period of 56.56 min. Although this period is close to
twice the exposure time of the LC data, we consider it unlikely that
this period is an artifact in the data. Since we can phase the SC data
on a common ephemeris it strongly points to the fact that the 56.56
min period is a signature of the orbital period. As this period is
below the short period minimum of hydrogen accreting CVs, {\src} may
be a helium-rich cataclysmic variable. The {\kep} light curve of
{\src} provides a unique insight to the accretion history of a
helium-rich CV over a 3 year timescale. Phase resolved spectroscopic
observations of {\src} are required when the system is in a bright
optical state to verify that the 56.56 min period is the orbital
period.

\section{Acknowledgments}

{\kep} was selected as the 10th mission of the Discovery Program.
Funding for this mission is provided by NASA, Science Mission
Directorate.  The {\kep} data presented in this paper were obtained
from the Multimission Archive at the Space Telescope Science Institute
(MAST).  STScI is operated by the Association of Universities for
Research in Astronomy, Inc., under NASA contract NAS5 26555. Support
for MAST for non HST data is provided by the NASA Office of Space
Science via grant NAG5 7584 and by other grants and contracts.  We
ackowledge support via NASA's {\kep} Grants 10-KEPLER10-0013 and
11-KEPLER11-0038. SH, SS, and DG wish to thank Carolyn Heffner and
Jean Mueller for their help and expertise at the 200$^{''}$ Hale
Telescope during our observations. Armagh Observatory is supported by
the Northern Ireland Government through the Dept of Culture Arts and
Leisure.

\vspace{ 8mm}

\end{document}